# Self-organization of supercooled liquids confined inside nano-porous materials

V. Teboul

Laboratoire POMA, UMR CNRS 6136

Université d'angers, 2 Bd Lavoisier 49045 Angers, France

Fax: +33241735216

E-mail: victor.teboul@univ-angers.fr

February 23$^{th}$ 2007

**Abstract:** Large scale molecular dynamics simulations are used to investigate the structural and dynamical modifications of supercooled water when confined inside an hydrophilic nanopore. We then investigate the evolution of the auto-organization of the most and the least mobile molecules (dynamical heterogeneity and string-like cooperative motions) when supercooled water is confined. Our calculations use the recent TIP5P intermolecular potential for water. We observe a strong slowing down of the dynamical properties when the liquid is confined, although the liquid structure is found to remain unchanged when corrected from the pore geometry. We then study cooperative motions inside supercooled confined water in comparison with bulk water. We observe strong modifications of the cooperative motions when the liquid is confined. We observe that dynamical heterogeneities and the associated correlation lengths are strongly increased as well as string-like motions in the confined liquid. This result, which is in opposition with the expected limitation of the correlation length by the confinement procedure, may explain (or be explained by) the slowing down of the dynamics. However the comparison of the dynamical heterogeneities at constant diffusion coefficient shows that the slowing down of the dynamics is not sufficient to explain the increase of the correlation lengths.

**Keywords:** heterogeneity, confinement, nanopore, glass transition, water, cooperativity.

## 1 Introduction

The microscopic dynamic behavior of water near hydrophilic surfaces is of fundamental interest for many biological and geological processes. Water is also of fundamental interest due to its many unique properties compared to other liquids [1]. Water is one of the most 'fragile' liquids in Angell's classification [2] of glass-formers. The comprehension of supercooled water is then important for the understanding of fragility. The existence of increased cooperative motion in supercooled glass forming liquids was long speculated [3,4] in order to explain the dramatic increase of their viscosity and various time scales with temperature decrease. Dynamical heterogeneities and associate cooperative behavior have been reported in various glass formers experimentally [5-9] near the glass transition temperature or with molecular dynamics (MD) simulations well above this temperature [5-9] and also in the glassy state [10]. Some of these heterogeneities were found to move string-like [11]. From MD simulations, dynamical heterogeneities are usually characterized by the aggregation of the most mobile molecules and also the aggregation of the least mobile molecules [5-9]. These two different aggregations and string-like cooperative motions have been observed previously in bulk supercooled water [12-15]. When a supercooled liquid is confined into a pore a few nanometers across, the cooperative motions cannot extend beyond the pore diameter. If the dynamics of the liquid is related as expected in supercooled liquids to the correlation length of cooperative motions, one can then expect an important modification of the dynamics with confinement when the pore diameter is smaller than this correlation length. Confinement appeared then as an indirect method to evaluate the correlation length of cooperative motions. In the simplest scenarios, one expects from finite size effects a cut-off on the slowing down of the dynamics as temperature is varied, when the typical size of the heterogeneities in the bulk would become larger than the diameter of the pore. Indeed an important modification of the dynamics is observed experimentally and with MD simulations when supercooled liquids are confined into nanopores [16-27]. Depending on the respective natures of the confining and confined medium, the relaxation rates are increased or decreased in the confined liquid. In some cases an acceleration of the dynamics has been observed experimentally [19], while in other cases a slow down has been observed experimentally [16-18]. A slow down has also been observed with MD simulations [20-27] and is predicted by mode coupling theory (MCT) calculations with particular confinement procedures [28-30].

In this paper, we study the modification of the dynamical heterogeneity observed in bulk supercooled water, when it is confined inside a pore a few nanometer across. We use a confinement procedure that eliminates the density and structural modification with confinement and minimize surface effects [20-25]. We then study the modification of the dynamics and the modification of the dynamical heterogeneities with confinement, with the aim of finding a possible relation between these effects. We observe a considerable increase in the relaxation times and an important increase of the dynamical heterogeneities. This article is organized as follows: in section 2 we describe the simulation procedure, in section 3 we discuss the different results and section 4 is the conclusion.

## 2 Calculation

The present simulations were carried out for a system of 507 water molecules (507 O + 1014 H). We have used the TIP5P potential [31-32], which has been found to reproduce well the structural and dynamical properties of water [31-34]. We used the densities calculated in ref. [31] and corresponding to a constant atmospheric pressure for this potential. The equations of motions were integrated with a Gear algorithm using the quaternions method [35]. The time step has been chosen equal to $10^{-15}$ s. The reaction field method [35] has been used to take into account long-range electrostatic interactions in the same conditions than described in ref.[31-32]. In order to minimize the effects due to the interactions with the surface of the pore, we use a particular confinement procedure here. Following references [20-23] the pore is made of vitreous water in purely elastic interactions with the liquid. We use the following procedure: The simulation begins with a parallelepipedic box of 25 Å x 25 Å x 50 Å that is equilibrated at the temperature of study. The liquid is then aged during 20 ns in order to insure stabilization. Then, molecules whose center of mass is positioned outside a pore of radius R= 8 Å are stopped instantaneously thus creating a wall for the other molecules which has the same structure than the bulk supercooled liquid. With this procedure the pore has the structure of the bulk liquid and one can expect that there is no structural modification of the liquid with confinement. We will check this point however in the next section. The liquid density is not modified by this procedure. Because the interactions with the wall are purely elastic, there is also no modification of the velocity distribution even in the vicinity of the wall. The confined liquid is then aged a second time during 20 ns before any investigation. Positions and velocities are then recorded and treated to obtain the statistical data.

In order to be able to compare the radial distribution function g(r) inside the pore and in the bulk liquid, the radial distribution function (RDF) of the confined liquid has to be corrected from the pore geometry. The radial distribution function

$$g(r) = \frac{V}{N^2} \left\langle \sum_{i,j=1} \delta(\mathbf{r} - \mathbf{r}_{ij}) \right\rangle$$

(1)

is computed in the bulk liquid simulation from the formula:

$$dN = \rho \cdot g(r) dV$$

(2)

with

$$dV = 4\pi r^2 dr = S(r) dr$$

(3)

dN is the number of molecules found inside the volume dV, S(r) is the surface of the sphere of radius r centered on the molecule *i* center of mass and $r=N/V$ is the density of the fluid. In the confined liquid the volume $dV = 4\pi r^2 dr = S(r) dr$ has to be replaced by

$$dV_{inside} = S_{in}(r,d) dr$$

(4)

where $S_{in}(r,d)$ is the surface of the part of the sphere of radius *r* centered on molecule *i* which is included inside the pore. In this calculation we limit ourselves to a virtual pore of radius R=6 Å in order not to consider molecules situated near the surface of the pore. The surface $S_{in}(r,d)$ depends on the distance *d* of molecule *i* to the pore axis. In a first step we calculate $S_{in}(r,d)$ versus the sphere radius *r* and the distance *d* of molecule *i* to the pore axis and the results are tabulated and recorded to be used in a second step in the different statistical calculations. The string-like mechanisms are investigated using the distinct part of the Van Hove correlation function:

$$G_d(\mathbf{r},t) = \frac{1}{N} \left\langle \sum_{i,j=1}^{N} \delta(\mathbf{r} + \mathbf{r}_i(0) - \mathbf{r}_j(t)) \right\rangle$$

(5)

In the figures we multiplied however $G_d(\mathbf{r},t)$ by the factor $\frac{V}{N}$ in order to obtain at time t=0 the radial distribution function

$$g(\mathbf{r}) = \frac{V}{N} G_d(\mathbf{r},0)$$

(6)

## 3. Results and discussion

This section is organized in three parts. In a first part we will show that although the structure remains unchanged, the dynamics is strongly slowed down when the liquid is confined. Then in a second part we will show that the correlation lengths of the structurally heterogeneous dynamic are not decreased but increased when water is confined. We will also show that in confined water the heterogeneous dynamics and string-like cooperative motions are strongly increased. Finally we will study the temporal evolution of dynamical heterogeneities in confined water and will compare it with that of bulk water.

The simplest way to study the structure of a liquid is to calculate the radial distribution function (RDF) g(r). This function g(r) (1) represents the probability of finding a water molecule at a distance *r* from another water molecule. On figure 1a we show the RDF g(r) between centers of mass of the water molecules in bulk and confined water at a temperature of 250 K. The radial distribution function in confined water is corrected here from the geometry of the pore, as explained in part II, in order to be compared directly with the radial distribution of the bulk liquid. We limited ourselves here to the structure inside a cylinder of 6 Å radius (in the center of the pore) so as not to consider possible perturbations in the vicinity of the wall. We observe on figure 1 that the two distributions (bulk and confined RDF) are roughly identical. This result shows that the radial structure is the same one in the two liquids. In order to study the angular distribution modification in the confined liquid we show in Figure 1b and 1c the RDF between hydrogen atoms corresponding to different molecules and the RDF between oxygen and hydrogen atoms. We observe on these figures a slight modification of these distributions. In Figure 1b the first and third peaks are slightly increased. And in Figure 1c the first peak is also slightly increased in the confined liquid. The structure thus appears roughly unchanged with the particular confinement procedure that we have used. This result is in agreement with previous works on confined Lennard-Jones liquids [20-23].

Figure 2 shows the mean square displacement (MSD) of the center of masses of the water molecules. The MSD are calculated in the direction of the pore axis $<z^2(t)>$ in the confined liquid. In the bulk due to the isotropy in displacement we have

$$<z^2(t)> = \frac{<r^2(t)>}{3}$$

(7)

We have then plotted for comparison the function $<r^2(t)>/3$ in bulk water at the same temperature of 250 K and at a lower temperature (240 K). The mean square displacement $<z^2(t)>$ is also displayed here for various distances from the center of the pore. Only are considered in this calculation the water molecules which remain between the two cylinders of radii n.R/4 and (n+1).R/4 over the time *t*, where R = 8 Å is the pore radius, and n is an integer value which varies between 0 and 3. We observe in figure 2 a strong slowing down of the dynamics inside the pore. The mean square displacement (MSD) is smaller near the wall and increases when we approach the center of the pore. However the mean square displacement inside the pore is smaller than the mean square displacement of the bulk liquid for each mean position simulated here. This result is in agreement with measurements in various confining media [16-18] and with previous MD simulations of supercooled water confined in hydrophilic mesoporous silica matrix [26-27]. A strong slowing down of the dynamics under confinement has also been observed in various supercooled liquids [20-25]. It has been found in previous simulations [24-27] of different liquids including water in hydrophilic pores [26-27] that the mean square displacement is roughly constant in the center of the pore and increase sharply near the pore surface. In our simulations we have also observed this result. However in order to improve Figure 2 clarity only one of the MSD of the two centrals layers which lead roughly to the same plot is displayed. We observe in Figure 2 that even the MSD of the third layer is not very different from the MSD in the pore center. This result will allow us to study the liquid in the center of the pore independently from the remaining liquid, allowing a direct comparison with the bulk liquid. We observe on figure 2 the three time regimes characteristic of supercooled liquids: The ballistic regime for times shorter than 0.3 ps, then the plateau regime between 0.3 ps and 20 ps and finally the diffusive regime for times higher than 20 ps. Near the wall, the plateau time regime appears however much larger. This large plateau time regime may be related with the large amount of heterogeneity that we have observed near the wall. Figure 2 shows that in the diffusive time regime the mean square displacement in the center of the pore is smaller than in the bulk at the same temperature of 250K. However Figure 2 shows that the MSD and then the diffusion coefficient in the pore at a temperature of 250K is the same than in the bulk at a temperature of 240K. We will use this result to compare the heterogeneities in the pore and in the bulk at constant diffusion coefficient.

With the aim of connecting this dynamical behavior to a modification of the heterogeneous dynamics we will now study dynamical aggregations in the center of the pore as well as string-like cooperative motions. These aggregations and string-like motions were observed previously in bulk supercooled water [12-15] in the same conditions of density and temperature. Dynamical heterogeneity is usually characterized by an aggregation of the most mobile molecules and an aggregation of the least mobile molecules, while the molecules of intermediate mobility [36] do not present any particular aggregation [5-9].

In figure 3 we display the functions

$$A^{+,t}(r) = \frac{g_{m-m,t}(r)}{g(r)} - 1$$

(8)

and

$$A^{-,t}(r) = \frac{g_{lm-lm,t}(r)}{g(r)} - 1$$

(9)

versus distance $r$, for confined and bulk supercooled water. In these formula $g_{m-m,t}(r)$ stands for the radial distribution function (RDF) between centers of masses of the 6% most mobile molecules and $g_{lm-lm,t}(r)$ the RDF between the 6% least mobile molecules. In order to eliminate the infinite values that appear when $g(r)=0$, we define here $A^{+,t}(r)=A^{-,t}(r)=0$ when $g(r)<0.1$. The mobility $\mu_i(t,t_0)$ which is used in the selection of the most or the least mobile molecules is here defined as the displacement of the molecule $i$ from time $t_0$ during a characteristic time $t$ [36].

$$\mu_i(t,t_0) = |\mathbf{r}_i(t+t_0) - \mathbf{r}_i(t_0)|$$

(10)

Because the functions $A^{+/-,t}(r)$ depend on the choice of this $t$ value, $t$ is indicated as indices in the formula. We chose here $t=t^+$ for $A^{+,t}(r)$ and $t=t^-$ for $A^{-,t}(r)$. Where $t^+$ is defined as the time that leads to the maximum of the integral:

$$I^+(t) = \frac{N}{V} \int_0^{R_{cutoff}} A^{+,t}(r) 4\pi r^2 dr$$

(11)

and similarly $t^-$ is defined as the time that leads to the maximum of the integral:

$$I^-(t) = \frac{N}{V} \int_0^{R_{cutoff}} A^{-,t}(r) 4\pi r^2 dr$$

(12)

Then $t^+$ and $t^-$ correspond to the characteristic times of the aggregation of the most and the least mobile molecules respectively. The $t^+$ and $t^-$ values at the temperatures of study are listed in Table I. In the literature the time corresponding to the maximum of the non Gaussian parameter (usually called $t^*$) is often used for the selection of the mobilities. However in previous works [15,37-38] we have shown that $t^+=t^*$ in supercooled water and liquid silica but that $t^-$ is usually larger than $t^*$. This coincidence of the times $t^*$ and $t^+$ appears due to the predominance of the large mobilities (then large mean square displacements) values in the non-Gaussian parameter [15,37-38]. When $g_{m-m,t}(r)=g(r)$ the function $A^{+,t}(r)$ is equal to zero, and this function $A^{+,t}(r)$ increases as $g_{m-m,t}(r)$ increases relatively to $g(r)$. And function $A^{-,t}(r)$ behaves in the same way. Functions $A^{+/-,t}(r)$ may then be used to quantify the aggregation of the most or the least mobile molecules. In other words, functions $A^{+/-,t}(r)$ represent the increase in probability of finding a mobile or non mobile molecule in the vicinity of another mobile or non mobile molecule. This increase in probability characterizes the presence of an aggregation of the mobile or non-mobile molecules respectively [11].

Figure 3a shows the function $A^{+,t+}(r)$ in bulk supercooled water at a temperature of 250K together with the same function in confined supercooled water (center of the pore) and in bulk supercooled water at a temperature (240K) corresponding to the same diffusion coefficient than in the center of the pore. We observe in figure 3a and b that the functions $A^{+/-,t}(r)$ are different from zero in the bulk and in the confined liquid. This result shows the presence of dynamical heterogeneity inside the bulk and confined liquid at a temperature of 250 K. We observe that the function $A^{+,t+}(r)$ is larger in the pore than in the bulk at the same temperature. The aggregation of the most mobile molecules is then increased by the confinement. Moreover we see in figure 3a that $A^{+,t+}(r)$ is larger in the pore at a temperature of 250 K than in the bulk at a temperature of 240K which corresponds to the same diffusion coefficient. Then for the same diffusion coefficient, in the pore the dynamic heterogeneity are larger than in the bulk. We define the size $\zeta_1^{+/-}$ of the aggregation by $A^{+/-,t^{+/-}}(\zeta_1^{+/-}) = 0.1$. With this definition we consider that there is a specific aggregation of the most mobile molecules only when the RDF between the most mobile molecules is at least 10% larger than the RDF between mean molecules. Another possible definition for the size of the aggregation follows from an exponential fit of the function $A^{+/-,t^{+/-}}(r)$: $A^{+/-,t^{+/-}}(r) = A_0^{+/-} e^{-r/\zeta_0^{+/-}}$. The values of $\zeta_1^{+/-}$, $A_0^{+/-}$ and $\zeta_0^{+/-}$ obtained, at the temperatures of study, from these formula are listed in Table I. We observe an average size of the aggregation $\zeta_1^+ = 11.9$ Å for the most mobile molecules in the pore center compared with $\zeta_1^+ = 9.2$ Å in the bulk liquid at 250K and $\zeta_1^+ = 10.1$ Å in the bulk liquid at 240K. This result is surprising for us as we expect confinement to hinder cooperativity for distances larger than the pore radius R=8 Å and then to decrease the dynamical heterogeneity at this temperature.

The situation is even more drastic for molecules of low mobility, as displayed in Table I and figure 3b. For the least mobile molecules, we observe an average size of the aggregation $\zeta_1^- = 27.5$ Å in the pore center compared with $\zeta_1^- = 9$ Å in the bulk liquid at 250K and $\zeta_1^- = 10.8$ Å in the bulk liquid at 240K. The aggregation size is for molecules of low mobility even larger than the pore diameter. This increase of the aggregation size can thus correspond only to an increase in the direction of the pore axis. Moreover, we observe that in the bulk liquid the size of the aggregation is roughly the same for the most and the least mobile molecules i.e. $\zeta_1^+ = \zeta_1^-$ in the bulk (at 240K: $\zeta_1^+ = 10.1$ Å and $\zeta_1^- = 10.8$ Å; and at 250K: $\zeta_1^+ = 9.2$ Å and $\zeta_1^- = 9$ Å); but this situation is broken in the confined liquid. In the pore center we observe a very important difference between the size of the two kinds of aggregations (we have $\zeta_1^+ = 11.9$ Å and $\zeta_1^- = 27.5$ Å).

We will now study the time evolution and characteristic times of the heterogeneity in bulk and confined supercooled water. For this purpose we will use the functions $I^{+/-}(t)$ defined previously as the integral of the functions $A^{+/-,t}(r)$. This function measures the aggregation of the most or the least mobile molecules for a characteristic time t and as in previous papers [15,37-38] we will call it the intensity of the aggregation at time $t$ for text clarity. Figure 4 shows the intensity $I^{+/-}(t)$ of the dynamical aggregation versus time for bulk and confined water. Figure 4 shows that the intensities of the aggregations are weak in the ballistic short time regime then increase with time, reach a maximum for a characteristic time $t=t^{+/-}$ located at the end of the plateau time regime and decrease as the diffusion regime is reached (see the MSD in Figure 2 for the time scales corresponding to these regimes). Figure 4 and Table I show that the characteristic time corresponding to the least mobile molecules aggregation is always larger than the characteristic time of the most mobile molecules aggregation. When the liquid is confined Figure 4 shows that the characteristic times and the intensity of the aggregation increase both for the most mobile and the least mobile molecules. We observe a particularly huge increase of the characteristic times associated to the least mobile molecules aggregation. Similarly the intensity increase is much more important for the least mobile molecules than for the most mobiles. Moreover this effect is attenuated in the figure due to the relatively short cutoff used in the calculation of the functions $I^{+/-}(t)$. Because the correlation lengths follow the relation $\zeta_1^+ < \zeta_1^-$, the cutoff of the function $A^{+,t}(r)$ leads to a less important decrease of the integral $I^+(t)$ than the cutoff of the function $A^{-,t}(r)$. We then observe in Figure 4 a dissymmetry in the increase of the correlation lengths, characteristic times and intensities of the aggregations of the most mobile molecules in comparison with the aggregation of the least mobile molecules.

Figure 5 shows the distinct Van Hove correlation function $G_{m-m,t+}(r,t)$ for the most mobile molecules (selected using the mobility at time $t^+=t^*$) in bulk and confined supercooled water. Two times are considered in $G_{m-m,t+}(r,t)$: $t=0$ and $t=t^+=t^*$. The function $G_{m-m,t+}(r,t)$ represents the probability of finding a mobile water molecule at time $t+t_0$ a distance $r$ apart from the position of another mobile molecule at time $t_0$. The increase in this function in $r=0$ for $t=t^+$ shows the increase in the probability of finding a mobile molecule with the position occupied previously by another mobile molecule. This increase is characteristic [11] of the presence of string-like motions in supercooled liquids. The Van Hove correlation function is corrected here, as explained previously, from the geometry of the pore. With the normalization used in Figure 5 at time $t=0$ the displayed Van Hove is simply the radial distribution function (6). We observe for $t=0$ an increase in the radial distribution function inside the pore compared to the same function calculated in the bulk liquid. This increase corresponds to the increase observed in the functions $A^{+,t+}(r)$ and shows the increase of the dynamic aggregation of the most mobiles molecules when water is confined. For $t=t^+$ we observe on the same figure the increase in $G_{m-m,t+}(r,t^+)$ for $r=0$. These increases, as explained above, are the signature of string-like motions in the liquid. We observe that the increase is much more important in confined water than in bulk water, at the same temperature. We thus observe an important increase of string-like motions when the liquid is confined.

## 4. Conclusion

We have studied the evolution of dynamical heterogeneity and string-like cooperative motions when supercooled water is confined into a pore a few nanometers across. The modification of the dynamics with confinement is usually seen as an indirect probe of the correlation length limitation by the pore radius, leading to an expected acceleration of the dynamics instead of the slowing down that is observed in most simulations and experiments. However we have found an increase of the correlation lengths when water is confined inside the pore, instead of the expected decrease. And this result may shed new light on the relationship between the modification of the dynamical properties and the auto-organization of the most and of the least mobile molecules (the so called dynamical heterogeneities). We have found that, at constant temperature, the dynamical heterogeneities increase when water is confined inside the pore. Using then a constant diffusion coefficient instead of a constant temperature we have also observed an increase of the dynamical heterogeneity with confinement. This result shows that the increase of the cooperativity is not a simple consequence of the slowing down of the dynamics.

**Figures:**

**Figure 1a:**

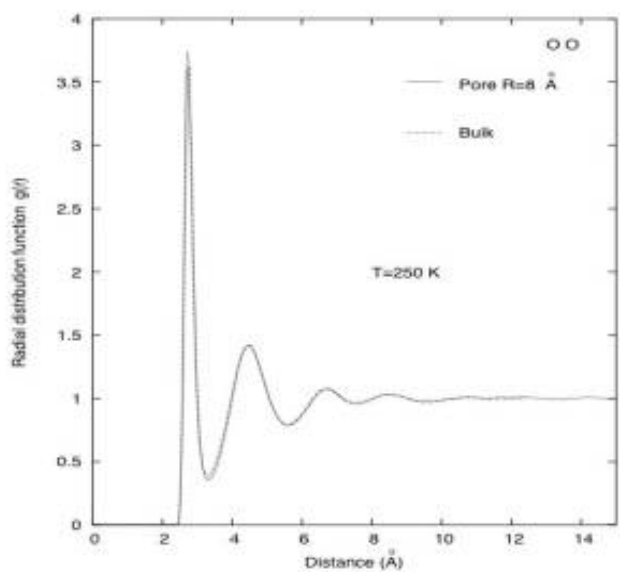

Radial distribution function between centers of masse, in confined water (dotted line) compared with that of water 'bulk' (continuous line). The temperature is 250K and the pore has a radius of 8 Å. The radial distribution function in the pore is numerically corrected from the geometry of the pore in order to allow a direct comparison between bulk and confined water.

**Figure 1b:**

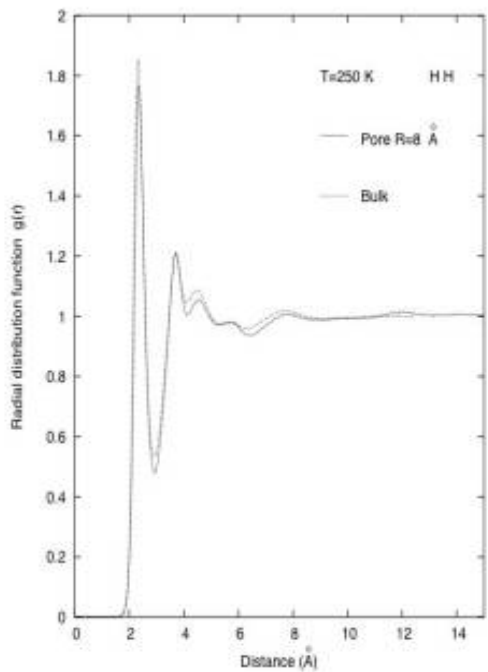

As in figure 1a, but between the hydrogen atoms of distinct water molecules.

**Figure 1c:**

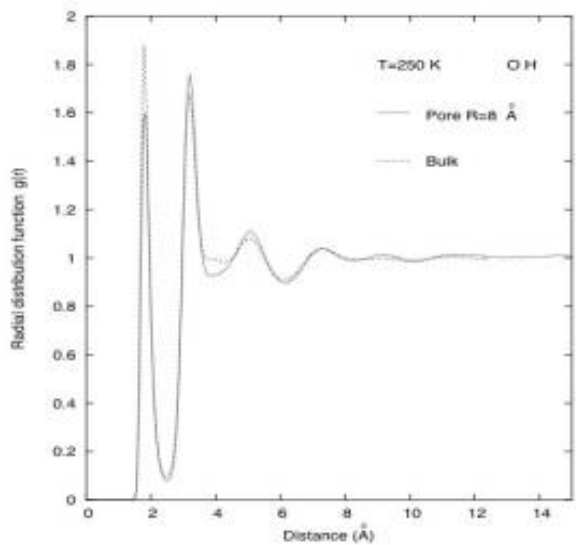

As in figure 1a, but between the oxygen and hydrogen atoms of distinct water molecules.

**Figure 2:**

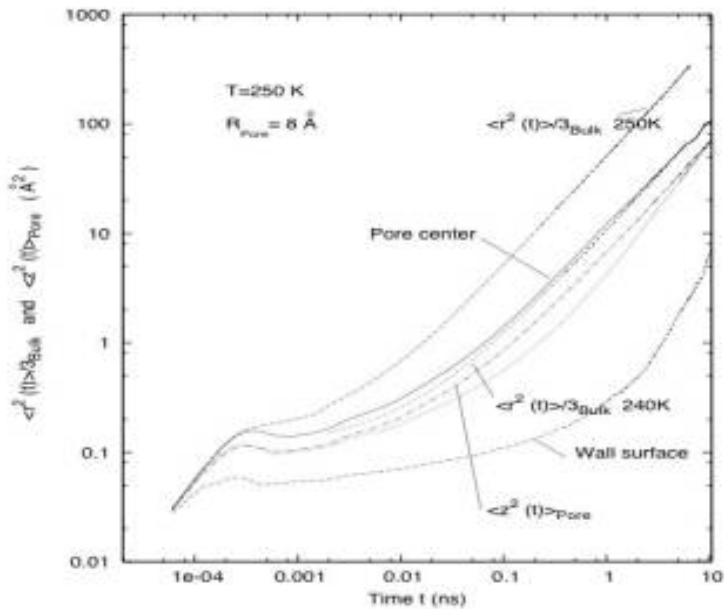

Mean square displacement $<z^2(t)>$ in the direction of the axis of the pore at a temperature of 250K compared to the mean square displacement of bulk water in one direction (in the bulk $<z^2(t)>=<r^2(t)>/3$) at two different temperatures: 240K and 250K. From top to bottom: a) Dashed line: $<z^2(t)>$ in the bulk at a temperature of 250K.
b) Continuous line: $<z^2(t)>$ in the pore centre.
c) Short dashed line: $<z^2(t)>$ in the bulk at a temperature of 240K.
d) Point dashed line: $<z^2(t)>$ mean value for the whole pore.
e) Dotted line: $<z^2(t)>$ between the centre and the wall of the pore.
f) Dashed line: $<z^2(t)>$ near the wall of the pore.

**Figure 3a:**

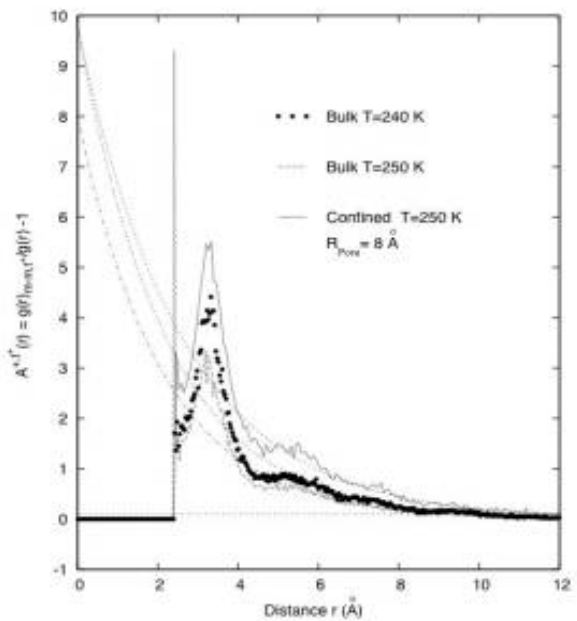

Function $A^{+J}(r) = \frac{g_{m-m,t}(r)}{g(r)} - 1$ for supercooled confined water and bulk water. In this formula $g_{m-m,t}(r)$ stands for the radial distribution function between the most mobile molecules. For this calculation the characteristic time used for the mobility has been chosen to correspond to the maximum of the function $I^+(t) = \frac{N}{V} \int_0^{R_{cutoff}} A^{+J}(r) 4\pi r^2 dr$. This characteristic time was found not to depend on the cutoff value $R_{cutoff}$, which is introduced here in order to limit noise fluctuations at large $r$, and to be able to compare results from different simulation box sizes.

**Figure 3b:**

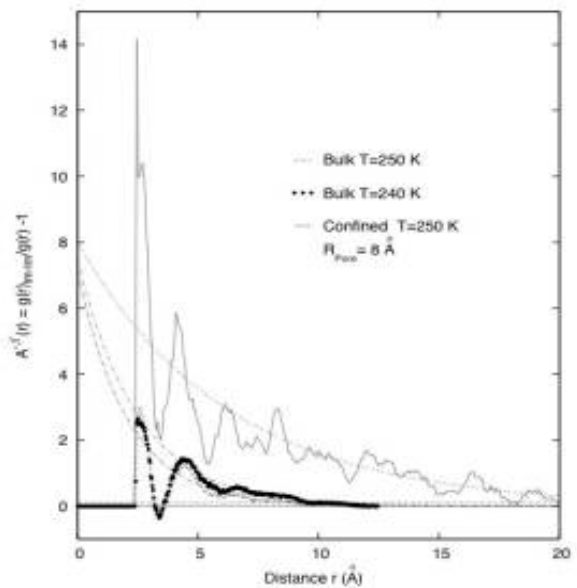

Function $A^{-,t}(r) = \dfrac{g_{lm-lm,t}(r)}{g(r)} - 1$ for supercooled confined water and bulk water. In this formula $g_{lm-lm,t}(r)$ stands for the radial distribution function between the least mobile molecules. For this calculation the characteristic time used for the mobility has been chosen to correspond to the maximum of the function $\Gamma(t) = \dfrac{N}{V}\int_0^{R_{cutoff}} A^{-,t}(r) 4\pi r^2 dr$. This characteristic time was found not to depend on the cutoff value $R_{cutoff}$, which is introduced here in order to limit noise fluctuations at large $r$, and to be able to compare results from different simulation box sizes. The function $A^{-,t}(r)$ in confined water has been slightly smoothed in this figure.

**Figure 3c:**

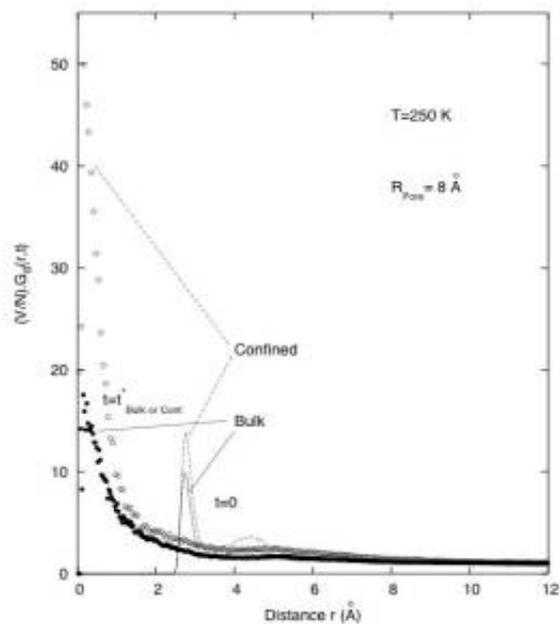

Distinct Van Hove correlation function for t=t* and t=0 in confined supercooled water compared to the same correlation function in bulk water.

**Figure 4:**

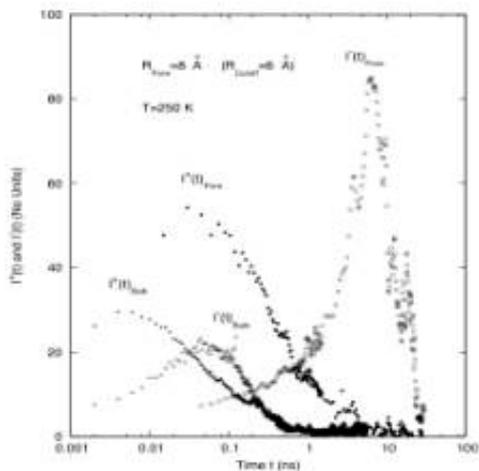

Functions $I^+(t) = \frac{N}{V}\int_0^{R_{cutoff}} A^{+,t}(r) 4\pi r^2 dr$ and $I^-(t) = \frac{N}{V}\int_0^{R_{cutoff}} A^{-,t}(r) 4\pi r^2 dr$ versus time t, for supercooled confined water and bulk water. $R_{cutoff}$ is introduced here in order to limit noise fluctuations at large $r$, and to be able to compare results from different simulation box sizes. The main effect of $R_{cutoff}$ is to decrease the value of $I^{+/-}(t)$ [38]. A decrease of $R_{cutoff}$ was found to decrease $I^-(t)$ more than $I^+(t)$. The characteristic time $t^{+/-}$ was however found unchanged by the modification of $R_{cutoff}$.

## Table I

|  | Characteristic time $t^{+/-}$ (ps) | Characteristic value $A_0^{+/-}$ (no units) | Characteristic size $\xi_0^{+/-}$ (Å) | Characteristic size $\xi_1^{+/-}$ (Å) |
|---|---|---|---|---|
| 250 K Bulk most mobiles (+) | 8 | 8 | 2.1 | 9.2 |
| 240K Bulk most mobiles (+) | $2.6 \cdot 10^1$ | 9.8 | 2.2 | 10.1 |
| 250K Confined most mobiles (+) | $4. \cdot 10^1$ | 9.8 | 2.6 | 11.9 |
| 250K Bulk least mobiles (-) | $6. \cdot 10^1$ | 7.1 | 2.1 | 9. |
| 240K Bulk least mobiles (-) | $5. \cdot 10^2$ | 7.5 | 2.5 | 10.8 |
| 250K Confined least mobiles (-) | $7. \cdot 10^3$ | 8.0 | 6.2 | 27.5 |

Characteristic times, sizes of the aggregations and maximum values of A(r,t) in bulk and confined water at different temperatures. Characteristic times $t^{+/-}$ are here defined as the times corresponding to the maximum of the functions $I^{+/-}(t)$. The correlation lengths $\xi_0$, $\xi_1$ and the maximum values $A_0^{+/-}$, are defined by the relations:

$$A^{+/-,t^{+/-}}(r) = A_0^{+/-} e^{-r/\xi_0^{+/-}} \quad \text{and} \quad A^{+/-,t^{+/-}}(\xi_1^{+/-}) = 0.1$$